\documentclass[doublecol]{epl2}

\title{Inverse Cotton-Mouton effect of the Vacuum and of atomic systems}

\author{C. Rizzo\inst{1,2,3} \and A. Dupays\inst{2,3} \and R. Battesti\inst{1} \and M. Fouch\'e\inst{2,3}
\and G.L.J.A. Rikken\inst{1}
}
\shortauthor{C. Rizzo \etal}
 \institute{
  \inst{1} Laboratoire National des Champs Magn\'etiques Intenses
(UPR 3228, CNRS-INSA-UJF-UPS), F-31400 Toulouse Cedex, France \\
  \inst{2} Universit\'e de Toulouse, UPS, Laboratoire Collisions Agr\'egats
R\'eactivit\'e, IRSAMC, F-31062 Toulouse, France\\
  \inst{3} CNRS, UMR
5589, F-31062 Toulouse, France\\
}
\pacs{nn.mm.xx}{First pacs description}
\pacs{nn.mm.xx}{Second pacs description}
\pacs{nn.mm.xx}{Third pacs description}

\abstract{In this letter we calculate the Inverse Cotton-Mouton
Effect (ICME) for the vacuum following the predictions of Quantum
ElectroDynamics. We compare the value of this effect for the
vacuum with the one expected for atomic systems. We finally show
that ICME could be measured for the first time for noble gases
using state-of-the-art laser systems and for the quantum vacuum with near-future laser facilities like ELI and HiPER, providing in particular a test
of the nonlinear behaviour of quantum vacuum
at intensities below the Schwinger limit of $4.5 \times 10^{33}$ W/m$^2$.}

\begin{document}

\maketitle

The advent of laser sources in the 1960s has opened the way to non-linear
optics thanks to the rapid increase in the light intensities which reached
$10^{19}$\,W/m$^2$ in the 1980s, and that can be nowadays as high as $10^{25}$\,W/m$^2$
\cite{Mourou}. Near-future laser facilities like the Extreme Light
Infrastructure (ELI) \cite{ELI} and the High Power laser Energy
Research system (HiPER) \cite{Hiper} should deliver $10^{29}$\,W/m$^2$ approaching
the Schwinger limit of
$4.5 \times 10^{33}$\,W/m$^2$ \cite{Mourou}. At this intensity optical nonlinearities of quantum vacuum
should be experimentally
accessible and quantum vacuum studies are one of the main motivations
to further increase laser intensity \cite{Marklund}.

In 1999 measurements of quantum electrodynamics processes in an intense electromagnetic wave, have been reported by Bamber {\it et al.} \cite{Bamber}. Nonlinear Compton scattering and electron-positron pair production have been observed in collisions between a laser beam of intensity up to $5\times10^{21}$\,W/m$^2$ and electrons of energy close to 50 GeV. The electric field strength of the laser in the electron rest frame corresponded to a few percents of the Schwinger limit.

Recently, an experiment coupling a very intense transverse pulsed magnetic field with an intense laser source has been performed \cite{Fouche}.
The goal was to detect a possible oscillation of photons into massive particles.
The maximum value of the pulsed magnetic field was about 10\,T over 0.36\,m, pulse duration was a few milliseconds. The laser source intensity was about
$10^{19}$\,W/m$^2$, corresponding to about 1500 J, over  5\,ns focussed on a spot of 100\,$\mu$m diameter.
These two pulsed facilities proved to work ideally together, opening new possibilities for
studies of non linear optics effects where a strong magnetic field and a powerful light source are necessary.
One of these effects is the inverse Cotton-Mouton effect (ICME in the following),
a non
linear optical effect that in principle exists in any
medium. In the presence of a transverse magnetic
field, a linearly polarized light induces a magnetization in the
medium in which it propagates \cite{Shen}. The optically induced
magnetization depends linearly on the transverse magnetic field
amplitude. ICME, as its name indicates, is related to the much
more studied Cotton-Mouton effect (CME in the following) i.e. the linear birefringence
induced by a transverse magnetic field \cite{RizzoRizzo} in a similar
fashion as the Faraday effect and the inverse Faraday effect are related \cite{Shen}.
ICME and CME can be explained as a mixing of four waves, two static
fields, and two photonic fields. The CME depends on the square of the amplitude of the
transverse magnetic field. To measure it, intense magnetic fields
are necessary. ICME depends on the transverse magnetic field amplitude, and to the light intensity. To measure such
an effect one needs to couple a powerful laser beam to an intense
magnetic field  transverse with respect to the light wave vector as in the experiment of ref. \cite{Fouche}.

As far as we know, experimental observations of ICME are
very rare. In Ref. \cite{Marmo} measurements of ICME in atomic gases are
cited \cite{Dabagyan,Movsesyan} for the case of resonant optical
pumping. Measurements of this kind of effect, called in ref. \cite{Shen} induced magnetization by resonant excitation, can be traced back to the sixties \cite{VanDerZiel}.

A measurement of the magnetization in a magnetically ordered crystal illuminated by a laser beam in the presence of a static magnetic field has been
reported in ref. \cite{Zon}. A linearly polarized beam from a
neodymium laser ($\lambda$=1,064 $\mu$m) with a pulse duration of
20 ns was focussed on a film of 10$\mu$m thickness of
(Lu,Bi)$_3$(Fe,Ga)$_5$O$_{12}$ immersed in a magnetic field
ranging from a few 10$^{-4}$ T to 3.10$^{-3}$ T. Measurements have
been conducted using laser energy between 4 and 20 mJ, beam spot
diameter was 1.3 mm, corresponding to intensity between $1.5 \times 10^{11}$ W/m$^2$ and $7.5 \times 10^{11}$ W/m$^2$.
The magnetization of the order of 10$^{-8}$ T
was measured by a planar three-turn coil on the surface of the
sample.
The measured magnetization did not depend on the laser polarization. The authors of ref. \cite{Zon} have called the phenomenon that they have observed ICME which is questionable since their static magnetic field is parallel to the direction of propagation of light. This kind of geometry is usually called Faraday configuration and it is associated in general to Faraday effects.

As far as we know a complete study of ICME for molecules does not
exist in literature. In ref. \cite{Marmo} one finds a
theoretical expression for the magnetization ${\bf M}_{\mathrm{ICM}}$
related to the ICME in the case of atoms. ${\bf M}_{\mathrm{ICM}}$ is
proportional to the elements of second hypermagnetizability tensor
$\eta_{\alpha\beta,\gamma\delta}$ on which also CME depends.

In this letter we calculate the inverse Cotton-Mouton effect for the quantum vacuum
following the predictions of Quantum ElectroDynamics. We
compare the value of this effect for the vacuum with the one
expected for atomic systems. We finally show that ICME could be
measured for the first time for noble gases using state-of-the-art laser systems and in the case of quantum vacuum with near-future laser facilities like ELI \cite{ELI} and HiPER  \cite{Hiper}, providing in particular a test of the nonlinear behaviour of quantum vacuum at intensities below the Schwinger limit.
\\

Optical non linearities in the propagation of light in vacuum have
been predicted since 1935 by the work of Euler and Heisenberg
\cite{EulerKockel,Heisenberg}. In particular vacuum in the presence of
a static magnetic field should behave as an uniaxial birefringent
crystal\cite{RikRiz}. This phenomenon is in all the aspects similar to what is
generally known as Cotton-Mouton effect\cite{RizzoRizzo}. The CME of quantum vacuum
has not yet been observed in spite of several experimental attempts (see \cite{Battesti} and references within).
In the 1936 paper by
Heisenberg and Euler \cite{Heisenberg} the complete study of the
phenomenon can be found, together with the general expression of
the non linear effective Lagrangian of the light-light
interaction.

The form of the effective lagrangian $L_{\mathrm{HE}}$ of the light-light
interaction is determined by the fact that the lagrangian has to
be relativistically invariant and therefore can only be a function of
the Lorentz invariants F, G:
\begin{equation}
F =  (\epsilon_{0}E^2 - {B^2 \over \mu_{0}})
    \label{2}
    \end{equation}
\begin{equation}
G =\sqrt{\epsilon_{0} \over \mu_{0}} ({\bf E} \cdot {\bf B})
    \label{3}
    \end{equation}
where $\epsilon_{0}$ is the vacuum permittivity, $\mu_{0}$ is
the vacuum permeability and $E$ and $B$ are the electromagnetic fields. Up to forth order in the fields, $L_{\mathrm{HE}}$
can be written as $L_{\mathrm{HE}} = L_{0} + L_{\mathrm{EK}}$ where $L_{0}$ is the
usual Maxwell's term and $ L_{\mathrm{EK}} $ is the first order non linear
term first calculated by Euler and Kockel \cite{EulerKockel}.

$L_{\mathrm{EK}}$ is valid in the approximation that the fields vary very
slowly over a length equal to the reduced electron Compton
wavelength $\mathchar'26\mkern-10mu\lambda = {\hbar \over m_{e}c}$
during a time ${{t_{e}}={\mathchar'26\mkern-10mu\lambda \over c}}$:
\begin{equation}
{\hbar \over m_{e}c} {|\nabla E (B)|} \ll E (B)
    \end{equation}
\begin{equation}
{\hbar \over m_{e}c^2} |{\partial E (B)\over \partial t} | \ll E
(B)
\end{equation}
with $\hbar$ the Planck constant divided by $2\pi$, $m_{e}$ the electron mass and $c$ the speed of light in vacuum.

Moreover $E$ and ${B \over \sqrt{\epsilon _{0}\mu _{0}}}$ have to
be smaller than the critical field $E_{\mathrm{cr}} = {m_e^2c^3 \over
e\hbar}$ i.e. $B \ll 4.4 \times 10^9$ T and $E \ll 1.3 \times
10^{18}$\,V/m. with $e$ the elementary charge. The laser intensity which corresponds to an electric field
associated to the light wave equal to $E_{\mathrm{cr}}$ is $4.5 \times 10^{33}$ W/m$^2$.
This intensity value is what is usually called the Schwinger limit.

$L_{\mathrm{HE}}$ can be written as
\begin{equation}
L_{\mathrm{HE}} = \frac{1}{2}F + a(F^2 + 7G^2)
\end{equation}
where $L_{0}= \frac{1}{2}F$ and $L_{\mathrm{EK}} = a(F^2 + 7G^2)$.
The value of $a$ given by Euler-Kockel\cite{EulerKockel} is:
\begin{equation}
a = {2\alpha^2 \hbar^3 \over 45 m_{e}^4 c^5}
\end{equation}
with $\alpha$ the fine structure constant. This corresponds to $a = 1.7 \times 10^{-30}$\,m$^3$/J.

We are interested in the magnetization ${\bf M} = {{\bf B} \over \mu_{0}} - {\bf H}$.
The field ${\bf H}$ can be obtained thanks to the
relations:
\begin{equation}\label{Acca}
{\bf H} = -{\partial L_{\mathrm{HE}} \over \partial {\bf B}}
\end{equation}
which gives:
\begin{equation}
{\bf H} = -{1 \over 2}{\partial F \over \partial {\bf B}} - {2a} F
{\partial F \over \partial {\bf B}} - {14a} G {\partial G \over
\partial {\bf B}}
\end{equation}
and finally:
\begin{equation}
{\bf H} = {{\bf B} \over \mu_{0}} + 4{a}{{\bf B} \over \mu_{0}}F -
{14a}\sqrt{{\epsilon_{0} \over \mu_{0}} }{\bf E}G.
\end{equation}

In the case of the propagation of an electromagnetic plane wave,
to which the fields $E_\omega$ and $B_\omega$ are associated, in
the presence of a static magnetic field $B_0$, one can write $B =
B_\omega + B_0$ and $E = E_\omega$, with
$\epsilon_{0} E_{\omega}^2 - {B_{\omega}^2 \over \mu_{0}}=0$ and
$\sqrt{\epsilon_{0} \over \mu_{0}} ({\bf E}_\omega \cdot
{\bf B}_\omega)=0$.

Finally one gets:
\begin{equation}
F = -{1 \over \mu_{0}}[B_{0}^2 + 2({\bf B}_{0}\cdot
{\bf B}_{\omega})]
\end{equation}
\begin{equation}
G = \sqrt{{\epsilon_{0} \over \mu_{0}}}({\bf E}_{\omega}\cdot
{\bf B}_{0}).
\end{equation}

The magnetization can be written as:
\begin{eqnarray}\nonumber
{\bf M} &= &4a{{\bf B}_{0}+{\bf B}_{\omega} \over
\mu_{0}^2}(B_{0}^2+2{\bf B}_{\omega}\cdot {\bf B}_{0}) \\
& + &14a{\epsilon_{0} \over
\mu_{0}}{\bf E}_{\omega}({\bf E}_{\omega}\cdot {\bf B}_{0}).
\end{eqnarray}

In the case of interest laser intensities are such that $B_\omega$, $E_\omega/c \gg B_0$, and the
magnetization corresponding to the ICME has to depend
linearly on the external magnetic field amplitude and quadratically on the electromagnetic fields
associated to light wave.
Extracting from the previous equation the terms of that type, one therefore obtains:
\begin{equation}
{\bf M}_{\mathrm{ICM}} = 14a{\epsilon_{0} \over
\mu_{0}}{\bf E}_{\omega}({\bf E}_{\omega}\cdot
{\bf B}_{0})+8a{\bf B}_{\omega}{{\bf B}_{\omega}\cdot {\bf B}_{0}
\over \mu_{0}^2}.
\end{equation}

Let's now recall that the square of the laser fields ${E}_{\omega}^2$ and ${B}_{\omega}^2$ are related to the laser intensity $I$ by:
\begin{equation}\label{E-I}
\epsilon_0 {E}_{\omega}^2 = {I \over c} = {{B}_{\omega}^2 \over \mu_0}.
\end{equation}

Two cases are possible (${\bf E}_{\omega} \| {\bf B}_{0}$,
${\bf B}_{\omega} \bot {\bf B}_{0}$) or (${\bf E}_{\omega} \bot
{\bf B}_{0}$, ${\bf B}_{\omega} \| {\bf B}_{0}$). In the first
case one gets:
\begin{equation}\label{mpar}
    {\bf M}_{\mathrm{ICM}\|} = 14a \epsilon_{0} E_{\omega}^2  {{\bf B}_{0} \over
\mu_{0}}= 14a {I \over c}  {{\bf B}_{0} \over
\mu_{0}}.
\end{equation}
In the second case one obtains:
\begin{equation}\label{mper}
    {\bf M}_{\mathrm{ICM}\bot} = 8a {B_{\omega}^2 \over \mu_{0}}  {{\bf B}_{0} \over
\mu_{0}} = 8a {I \over c} {{\bf B}_{0} \over
\mu_{0}}.
\end{equation}

In both cases ${\bf M}_{\mathrm{ICM}}$ is parallel to ${\bf B}_{0}$. It is worth to stress
that the fact that ${\bf M}_{\mathrm{ICM}\|}\neq {\bf M}_{\mathrm{ICM}\bot}$ confirms that under the effect
of an external magnetic field, vacuum should become non isotropic, and its magnetic susceptibility should
depend on light polarization. Actually, electric polarizability should also become non isotropic and
finally the index of refraction should depend on light polarization. This is the cause of
the CME of quantum vacuum \cite{RikRiz}.

As said before, an ICME set up consists of a powerful laser and of
an intense transverse magnetic field. The most powerful lasers are
usually pulsed, the same applies to magnetic field generation \cite{LNCMI-T}. A
set up coupling these two instruments have been recently realized
in the framework of the search for photon oscillations into
massive particles \cite{Fouche}. A $10^{19}$\,W/m$^2$ laser pulse
was focused in a vacuum
region where a transverse magnetic field of more than 10 T was present.

Let's take these numerical values to have a reasonable estimate of
the magnetization to be measured in the case of the ICME of the quantum
vacuum:
\begin{equation}\label{mparnum}
    {\bf M}_{\mathrm{ICM}\|} \approx 8 \times 10^{-18} \mathrm{T}
\end{equation}
and
\begin{equation}\label{mpernum}
    {\bf M}_{\mathrm{ICM}\bot} \approx 4.5 \times 10^{-18} \mathrm{T}
\end{equation}
where we have used the relation $\mu_0 \bf M$(A/m)= $\bf M$(T).

Measurements of a magnetization induced by a laser beam are also
performed in the framework of the inverse Faraday effect (IFE). A
circularly polarized laser beam creates in a medium a
magnetization proportional to the energy density associated to the
electromagnetic wave \cite{Shen}. This effect is related to the
Faraday effect as the ICME is related to the Cotton-Mouton effect.
In the case of IFE measurements sensitivity in magnetization of
the order of $10^{-10}$ T has been reached \cite{Kalugin}. The
same kind of sensitivity should be reached in the case of ICME.

The values in Eqs.\,(\ref{mparnum}) and (\ref{mpernum}) are still below the sensitivity reported in
\cite{Kalugin}, but new laser sources like the Extreme Light
Infrastructure (ELI) \cite{ELI} and the High Power laser Energy
Research system (HiPER) \cite{Hiper} are supposed to reach
intensities exceeding $10^{29}$ W/m$^2$ increasing the expected
ICME of vacuum at levels that should be detectable.
In particular, taking also advantage of progress in tranverse pulsed magnetic field
\cite{LNCMI-T} and using a field of at least 30 T, a laser intensity of $5 \times 10^{25}$ W/m$^2$,
well below the possibilities of new facilities, will be sufficient
to open up direct studies of quantum vacuum with powerful laser systems.\\

In the following, for the sake of comparison, let's calculate the expected ICME
in the case of atoms and in particular noble gases.

Our calculation of the magnetization corresponding to the ICME in atoms is based on the
Buckingam and Pople general theory of molecular polarizabilities in the presence
of a strong magnetic field \cite{BuckPople}. In the framework of this theory the atomic
magnetic moment can be written as:
\begin{equation}\label{Matom}
    {\bf \mu}_{\mathrm{at}}= -{ dU \over d{\bf B}}
\end{equation}
where $U$ is the atomic energy in a strong external magnetic field.
$U$ can be expanded in a power series of the electromagnetic fields \cite{RizzoRizzo}. The upper limit of validity of such an approximation is not discussed in literature, but the comparison between measurements and theoretical values obtained using this expansion indicates that it is certainly valid for magnetic fields of several Teslas (see e.g. ref. \cite{RizzoRizzo}). We will assume in the following that it is also valid for higher fields.

It is important to stress that, as shown in ref. \cite{ARizzoCRizzo}, in the case of CME the effect of the interaction of the magnetic field associated with the propagating wave with the atomic or molecular system is very small compared to the main effect induced by the electric field of the wave and is usually neglected. We will assume in the following that the same applies to ICME.
This is not the case for the quantum vacuum as clearly shown by Eqs.\,(\ref{mpar}) and (\ref{mper}).

We are looking for an atomic magnetic moment which depends linearly on the external magnetic field
and quadratically on the electric field. Because of Eq.\,\ref{Matom}, this kind of induced magnetic moment
can only be obtained by derivating the term of the $U$ series quadratic in the electric and magnetic fields:
\begin{equation}\label{etaterm}
    U_\eta= -{1 \over 4}\eta_{\alpha \beta, \gamma \delta}E_\alpha E_\beta B_\gamma B_\delta
\end{equation}
where $\eta$ is the second hypermagnetizability tensor, Einstein summation is assumed and $(\alpha,\beta,\gamma,\delta) = x,y,z$.

To obtain the magnetization ${\bf{M}}^{\mathrm{at}}$ we have to multiply the atomic magnetic moment ${\bf \mu}_{\mathrm{at}}$
by the atom density which for ideal gases is equal to $P/kT$, where $P$ is the gas pressure, $k$ the Boltzmann
constant, and $T$ the temperature. Taking into account the two possible cases as for the quantum vacuum, we finally obtain:
\begin{equation}\label{mparperat}
    {\bf M}^{\mathrm{at}}_{\mathrm{ICM}\|,\bot} = {1 \over 2}{P \over kT} \eta_{\|,\bot} E_{\omega}^2  {\bf B}_{0}
\end{equation}
where $\eta_{\parallel(\bot)}$ is the component of the $\eta$
tensor parallel (perpendicular) to ${\bf B}_{0}$. These two
components are related to the Cotton-Mouton effect in atoms
\cite{RizzoRizzo} since the magnetic induced birefringence $\Delta
n=n_\parallel-n_\bot$ is proportional to $(\eta_\parallel -
\eta_\bot)$.
In both cases ${\bf M}_{\mathrm{ICM}}$ is parallel or antiparallel to ${\bf B}_{0}$ depending
on the sign of the $\eta$ component.
Our theoretical result is equivalent to the one given in Ref.\,\cite{Marmo}.

Formula (\ref{mparperat}) can also be written as:
\begin{equation}\label{mperparat}
    {\bf M}^{\mathrm{at}}_{\mathrm{ICM}\|,\bot} = {1 \over 2}{P \over kT} \eta_{\|,\bot} Z_0 I  {\bf B}_{0}
\end{equation}
where $Z_0 = \sqrt{{\mu_0 \over \epsilon_0}}=377$\,$\Omega$ is the vacuum impedance.

To get a numerical estimation of ${\bf M}^{\mathrm{at}}_{\mathrm{ICM}}$ in Tesla units, let's write Eq.\,(\ref{mperparat}) as follows:
\begin{equation}\label{mperparatNum}
    {\bf M}^{\mathrm{at}}_{\mathrm{ICM}\|,\bot} \simeq 5.1\times 10^{-28}{P \over T} \eta_{\|,\bot} I {B}_{0}
\end{equation}
where $P$ is given in atm, $T$ in K, $\eta_{\|,\bot}$ in atomic units (au in the following), $I$ in W/m$^2$, ${B}_{0}$ in T, and the resulting ${\bf M}^{\mathrm{at}}_{\mathrm{ICM}\|,\bot}$ is also given in T.
Let's also recall that 1 $\eta$(au) is equal to $2.98425 \times 10^{-52}$ C$^2$m$^2$J$^{-1}$T$^{-2}$
\cite{RizzoRizzo}.

Theoretical values of $\eta_{\|,\bot}$
for noble gases can be found in Ref.\,\cite{Bishop}. In table \ref{tab.1} we summarize our results obtained assuming that
$P= 1$ atm, $T=300$ K, and that $I=10^{19}$ W/m$^2$, $B=10$ T like in Ref.\,\cite{Fouche}.

\begin{table} [h]
\caption{Expected values of ICME magnetization for noble gases for $P= 1$ atm, $T=300$ K, $I=10^{19}$ W/m$^2$, $B=10$ T.}
\label{tab.1}
\begin{center}
\begin{tabular}{lllll}
gas  & $\eta_\|$(au) & ${\bf M}^{\mathrm{at}}_{\mathrm{ICM}\|}$(T) & $\eta_\bot$(au) & ${\bf M}^{\mathrm{at}}_{\mathrm{ICM}\bot}$(T)\\
He & -1.213 & -2.1$\times 10^{-10}$ & -2.1668 & -3.8$\times 10^{-10}$ \\
Ne & -2.040 & -3.5$\times 10^{-10}$ & -4.254 & -7.4$\times 10^{-10}$\\
Ar & -18.84 & -3.2$\times 10^{-9}$ & -41.21 & -7.1$\times 10^{-9}$\\
Kr & -38.11 & -6.6$\times 10^{-9}$ & -86.72 & -1.5$\times 10^{-8}$\\
Xe & -83.10 & -1.4$\times 10^{-8}$ & -200.85 & -3.4$\times 10^{-8}$
\end{tabular}
\end{center}
\end{table}

Comparing results of table \ref{tab.1} with results for quantum vacuum given by Eqs.\,(\ref{mparnum}) and (\ref{mpernum}),
one obviously finds that the effect in gases is many orders of magnitude bigger than the one predicted for quantum vacuum.
On the other hand in the case of gases one cannot increase the laser intensity arbitrarily because of gases
ionization. Laser ionization of noble gases has been studied in Ref.\,\cite{Augst} at $\lambda = 1.053$ $\mu$m. A systematic scan of intensities from $10^{17}$ W/m$^2$ to $10^{20}$ W/m$^2$ was performed. Ionization appears at different intensities
depending on the noble gas. For Helium and Neon ionization begins around $10^{19}$ W/m$^2$, for Argon and Kripton
around $10^{18}$ W/m$^2$, and for Xenon around $10^{17}$ W/m$^2$. Ion production rate is of the order of a few tens of ions at the intensities given before for a gas pressure of a few $10^{-9}$ atm. The consequent ion current could somewhat perturb the ICME measurement.

Result shown in table \ref{tab.1} places Helium at the limit of which is detectable with existing facilities, and
it also shows that ICME of other noble gases like Ne and Ar could be observed for the first time.
It is also important to notice that ICME could allow to measure $\eta_\|$ and $\eta_\bot$ separately,
while CME gives only access to the difference of the two.

In conclusion, in this letter we show that both ICME of quantum vacuum and ICME of atomic species can be measured
using near-future or existing laser facilities opening the way to the observation of a new phenomenon in dilute matter.

%
%

\acknowledgments
We thank A. Ben-Amar Baranga and A. Rizzo for carefully reading the manuscript and very useful discussions.


\begin{thebibliography}{0}



\bibitem{Mourou} \Name{G.A. Mourou, T. Tajima \and S.V. Bulanov} \REVIEW{Rev. Mod. Phys.}{78}{2006}{309}.

\bibitem{ELI} see http://www.extreme-light-infrastructure.eu/

\bibitem{Hiper} see http://www.hiper-laser.org/

\bibitem{Marklund} \Name{M. Marklund \and J. Lundin} \REVIEW{Eur. Phys. J. D}{55}{2009}{319}.

\bibitem{Bamber} \Name{ C. Bamber, S.J. Boege, T. Koffas, T. Kotseroglou, A.C. Melissinos, D.D. Meyerhofer, D.A. Reis, W. Ragg, C. Bula, K.T. McDonald, E.J. Prebys, D.L. Burke, R.C. Field, G. Horton-Smith, J.E. Spencer, D. Waltz, S.C. Berridge, W.M. Bugg, K. Shmakov, \and A.M. Weidemann} \REVIEW{Phys. Rev. D}{60}{1999}{092004}.

\bibitem{Fouche} \Name{M. Fouch\'e, C. Robilliard, S. Faure, C. Rizzo, J. Mauchain, M. Nardone, R. Battesti, L. Martin, A.-M.Sautivet, J.-L. Paillard \and F. Amiranoff} \REVIEW{Phys. Rev. D}{78}{2008}{032013}.

\bibitem{Shen}
Y.R. Shen, {\it The Principles of Nonlinear Optics}, (John Wiley \& Sons, New York 1984) 1st ed., p. 53 and seq..

\bibitem{RizzoRizzo}
\Name{C. Rizzo, A. Rizzo \and D.M. Bishop} \REVIEW{Int. Rev.
Chem. Phys.}{16}{1997}{81}.

\bibitem{Marmo}
\Name{S.I. Marmo \and V.D. Ovsiannikov} \REVIEW{Phys. Lett.
A}{202}{1995}{201}.

\bibitem{Dabagyan}
\Name{A.A. Dabagyan, M.E. Movsesyan \and R.E. Movsesyan}
\REVIEW{Pis'ma Zh. Eksp. Teor. Fiz.}{29}{1979}{586}.

\bibitem{Movsesyan}
\Name{M.E. Movsesyan} \REVIEW{Izk. Akad. Nauk. SSSR. Ser.
Fiz.}{45}{1981}{2230}.

\bibitem{VanDerZiel} \Name{J.P. Van der Ziel \and N. Bloembergen} \REVIEW{Phys. Rev. A}{4}{1965}{1287}.

\bibitem{Zon}
\Name{B.A. Zon, V.Ya. Kupershmidt, G.V. Pakhomov \and T.T. Urazbaev} \REVIEW{JETP Lett.}{45}{1987}{272}.

\bibitem{EulerKockel}
\Name{H. Euler \and B. Kockel} \REVIEW{Naturwiss.}{23}{1935}{246}.

\bibitem{Battesti} \Name{R. Battesti, B. Pinto Da Souza, S. Batut, C. Robillard, G. Bailly, C. Michel, M. Nardone, L. Pinard, O. Portugall, G. Tr\'enec J.-M. Mackowski, G.L.J.A. Rikken, J. Vigu\'e \and C. Rizzo} \REVIEW{Eur. Phys. J. D}{46}{2008}{323}.

\bibitem{Heisenberg}
\Name{W. Heisenberg \and H. Euler} \REVIEW{Z. Phys.}{38}{1936}{714}.

\bibitem{RikRiz} \Name{G.L.J.A. Rikken \and C. Rizzo} \REVIEW{Phys. Rev. A}{63}{2001}{012107}.

\bibitem{LNCMI-T} http://www.toulouse.lncmi.cnrs.fr/?lang=en

\bibitem{Kalugin}  \Name{N.G. Kalugin \and G. Wagni\`ere}
\REVIEW{Quantum Semiclass. Opt.}{3}{2001}{S189}.

\bibitem{BuckPople} \Name{A.D. Buckingam \and J.A. Pople}
\REVIEW{Proc. Phys. Soc. B}{69}{1956}{1133}.

\bibitem{ARizzoCRizzo} \Name{A. Rizzo \and C. Rizzo}
\REVIEW{Mol. Phys.}{96}{1999}{973}.

\bibitem{Bishop} \Name{D.M. Bishop, \and S.M. Cybulski}
\REVIEW{Chem. Phys. Lett.}{211}{1993}{255}.

\bibitem{Augst} \Name{S. Augst, D.D. Meyerhofer, D. Strickland \and S.L. Chint} \REVIEW{J. Opt. Soc. Am. B}{8}{1991}{858}.


\end{thebibliography}
\end{document}